\documentclass[final]{eps}
\usepackage{graphicx}
\usepackage{times}
\usepackage{geometry}
\volumenumber{x}
\publishyear{2013}

\frompage{\pageref{firstpage}}
\topage{\pageref{finalpage}}

\shorttitle{K. Malek \lowercase{\textit{et al.}}: Dusty Universe viewed by AKARI far infrared detector.}

\title{Dusty Universe viewed by AKARI far infrared detector}
\author{K.\ Ma{\l}ek$^{1}$, A.\ Pollo$^{2,3}$, T.\ T.\ Takeuchi$^{1}$, E.~Giovannoli$^{4,5}$, V.~Buat$^{4}$, D.~Burgarella$^{4}$, M.~Malkan$^{6}$, and A.~Kurek$^3$ }
\affiliation{	$1$ Department of Particle and Astrophysical Science, Nagoya University, Furo-cho, Chikusa-ku, 464-8602 Nagoya, Japan\\
	$^2$ National Centre for Nuclear Research, ul. Ho\.za 69, 00-681 Warszawa, Poland  \\
	$^3$ The Astronomical Observatory of the Jagiellonian University, ul.\ Orla 171, 30-244 Krak\'{o}w, Poland\\ 
	$^4$ Laboratoire d'Astrophysique de Marseille, OAMP, Universit$\acute{e}$ Aix-Marseille, CNRS, 38 rue Fr$\acute{e}$d$\acute{e}$ric Joliot-Curie, 13388 Marseille, cedex 13, France\\
        $^5$ University of the Western Cape, Private Bag X17, 7535, Bellville, Cape Town, South Africa\\
	$^6$ Department of Physics and Astronomy, University of California, Los Angeles, CA 90024}
\received{xxxx xx, xxxx}\revised{xxxx xx, xxxx}\accepted{xxxx xx, xxxx}
\abstract{
We present the results of the analysis of multiwavelength Spectral Energy Distributions (SEDs) of far-infrared galaxies detected in the AKARI Deep Field-South (ADF--S) Survey. 
The analysis uses a carefully selected sample of 186 sources detected at the 90 $\mu$m AKARI band, identified as galaxies  with cross-identification in public catalogues. 
For sources without known spectroscopic redshifts, we estimate photometric redshifts after a test of two independent methods: one based on using mainly the optical -- mid infrared range, and one based on the whole range of ultraviolet  -- far infrared data.
We observe a vast improvement in the estimation of photometric redshifts when far infrared data are included, compared with an approach based mainly on the optical -- mid infrared range. 
We discuss the physical properties of our far-infrared-selected sample.   
We conclude that this sample consists mostly of rich in dust and young stars nearby galaxies, and, furthermore, that almost 25\% of these sources are (Ultra)Luminous Infrared Galaxies. 
Average SEDs normalized at 90 $\mu$m for normal galaxies (138 sources), LIRGs (30 sources), and ULIRGs (18 galaxies) a the significant shift in the peak wavelength of the dust emission, and an increasing ratio between their bolometric and dust luminosities which varies from 0.39 
to 0.73.
}

\begin{document}
\label{firstpage}
\maketitle
\copyrighttext{}

\section{Introduction}

Star formation (SF) history holds the key to understanding galaxy evolution, and - from a wider perspective - the nature of our Universe.
One of the obstacles to observing starburst regions in galaxies using optical telescopes is dust.
On the other hand, protostars form from dust clouds and molecular gas.
Most of the dust in galaxies is quite cool ($\sim$ 10-20~K), and its emission is visible only in the far infrared ({Glass}, 1999). 
A warm dust component might be observable during starburst activity (Phillips, 2005). 
The newly born, blue massive stars are surrounded by gas and dust, which obscure the most interesting regions and,  additionally, absorb a part of the ultraviolet (UV) light emitted by stars. 
Dust heated in this way re-emits the absorbed light in the infrared wavelengths, mostly in the far infrared.  
Even very careful observations in the UV and optical ranges of wavelengths cannot provide a detailed description of the SF processes in galaxies. 
The infrared (IR) emission, reflecting the dust-obscured SF activity of galaxies (Genzel \& Cesarsky, 2000), combined together with UV and optical data, can give full information about the star formation history and rate. 
Additionally, the ratio between the UV and far infrared (FIR) emission may serve as an indicator of the dust attenuation in galaxies (e.g., Buat et~al., 2005, Takeuchi et~al., 2005, Noll et~al., 2009).
The first all-sky survey at IR was performed by the satellite IRAS (\textit{The Infra-Red Astronomical Satellite}, Neugebauer et~al., 1984). 
IRAS mapped the sky at 12~$\mu$m, 15~$\mu$m, 60~$\mu$m, and 100~$\mu$m for 300 days and forever changed our view of the sky (Beichman, 1987). 
IRAS detected IR emission from about 350~000 astronomical sources, and the number of known astronomical objects went up by 70\%. 
A large fraction of the extragalactic sources detected in the far infrared were spiral galaxies, quasars (QSOs), Seyfert galaxies, and early type galaxies (Genzel \& Cesarsky, 2000), but also new classes of galaxies very bright in the IR, 
such as \textit{U}LIRGs (\textit{ultra-}luminous infrared galaxies), were found.  
IR satellite missions, focusing on selected areas, were followed by MSX (\textit{Midcourse Space Experiment}, e.g., Egan et al., 2003), ISO (\textit{Infrared Space Observatory},e.g., Verma et al., 2005; Genzel \& Cesarsky, 2000), and SST (\textit{The Spitzer Space Telescope}, e.g., Soifer et al., 2008).

After more than 20 years, a Japanese spacecraft, AKARI (\textit{akari} means \textit{warm-light} in Japanese), performed a new all-sky survey with a much higher angular resolution than IRAS ({Murakami} et~al., 2007). 
AKARI has also provided deeper surveys centered on the North and South Ecliptic Poles (e.g., Wada et~al., 2008, Takagi et~al., 2012, Matsuura et~al., 2011).  
In particular, with the Far-Infrared Surveyor (FIS, Kawada et~al., 2007), observations in the four FIR bands were possible (65 $\mu$m, 90 $\mu$m, 140 $\mu$m, and 160 $\mu$m).
Among the observed fields, the lowest Galactic cirrus emission density region near the South Ecliptic Pole was selected for {deep} observation, {to} provide the best FIR extragalactic image of the Universe.
This field is referred to as the AKARI Deep Field South (\mbox{ADF--S}). 
This survey is unique in having a continuous wavelength coverage with four photometric bands and mapping over a wide area (approximately 12 square degrees). 
In the ADF-S, 2~263 infrared sources were detected down to $\sim$ 20 mJy at the 90 $\mu$m band, and infrared colours for about 400 of these sources were measured.
The first analysis of this sample in terms of the nature and properties for 1000 \mbox{ADF--S} objects brighter than 0.301 Jy in the 90 $\mu$m band was presented by {{Ma{\l}ek} et~al. (2010)}.

In this work, we present a more sophisticated  approach to the analysis than the previous analysis ({Malek} et al., 2010) of the \mbox{ADF--S} sources in multiwavelength studies, based on a sample of 545 identified galaxies. 
In Section~\ref{data}, we present our data and sample selection. 
We discuss the spectroscopic redshifts distribution in Section~\ref{spectroredshifts}, and a new approach to the estimation of photometric redshifts based on the Le~PHARE and CIGALE codes in Section~\ref{photoredshifts}. 
Discussion of physical and statistical properties of the obtained SEDs is presented in  Section~\ref{fitting}. 
The basic properties of a sample of galaxies with known spectroscopic redshifts are shown in Section~\ref{result}. 
A discussion of (U)LIRGs properties found in our sample is presented in Section~\ref{resultAVSEDS}.
Section~\ref{conclsection} presents our conclusions. 

In all calculations in this paper we assume the flat model of the Universe, with $\mathrm{\Omega_{M}}$ = 0.3, $\mathrm{\Omega_{\Lambda}}$ = 0.7, and $\mathrm{H_{0}}$ = 70 km $\mathrm{s^{-1}Mpc^{-1}}$.

\section{Data}
\label{data}
The main aim of our work is to build a galaxy sample with high quality fluxes from the UV to the FIR using the \mbox{ADF--S} sample. 
Redshift information is also needed to obtain physical parameters from the SEDs. 

Our sample is drawn from the AKARI \mbox{ADF--S} catalog presented by {Ma{\l}ek} et~al. (2010), and published at the Centre de Donn\'{e}es astronomiques de Strasbourg, Strasbourg astronomical Data Center (http://cdsarc.u-strasbg.fr/viz-bin/qcat?J/A+A/514/A11).
The sample consists of the 545 \mbox{ADF--S} sources from the so-called 6$\sigma$ catalog (S$\rm{_{90}\mu m>}$0.0301Jy, which corresponds to the 6$\sigma$ detection level) 
measured by the AKARI FIS detector, for which the optical counterparts were found in public catalogues ({Ma{\l}ek} et~al., 2010).
Additional measurements, mostly from WISE (Wright et~al., 2010) and GALEX ({Dale} et~al., 2007), and further information from public databases (SIMBAD {http://simbad.u-strasbg.fr/simbad/}, NED {http://ned.ipac.caltech.edu/}, and IRSA {http://irsa.ipac.caltech.edu/}) were used in our present analysis (see also {Ma{\l}ek} et~al., 2013). 

\section{Spectroscopic redshifts distribution}
\label{spectroredshifts}

In total (data coming from {Ma{\l}ek} et~al., 2010; the dedicated spectroscopic measurements of selected \mbox{ADF--S} sources performed by {Sedgwick} et~al., 2011; new data from the NED and SIMBAD databases) 
we have 173 galaxies with known spectroscopic redshifts  ($z_{\mathrm{spectro}}$).
The mean value of the redshift in this sample is equal to $0.08 \pm 0.01$, with a median value of 0.06. 
This implies, {together with AKARI FIR measurements,} that our sample mainly consists of nearby FIR-bright galaxies. 

	\begin{figure}[t]
		\centerline{\includegraphics[scale=0.55,clip]{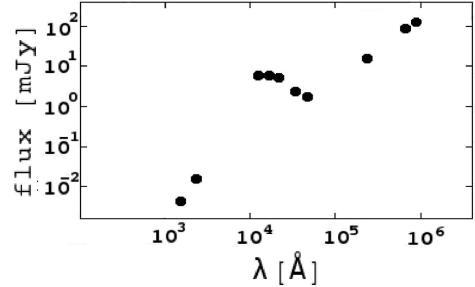}}
		\caption[]{ A typical example of a spectral coverage for a source (here: 2MASX J04421266-5355520) within our sample. 
		Each galaxy from our sample has  a measurement in the FIR and optical parts of its spectrum. 
		Nearly 50\% of galaxies have additional detections in the UV and MIR.}
		\label{6M}
	\end{figure}

Unfortunately, the majority of our sample with spectroscopic redshift information does not have enough photometric measurements to fit SED models with a high confidence. 
 {After a visual inspection of all the spectra}, we decided to apply a more stricter selection criterion, and for the next procedure we have chosen  only galaxies with  at least six photometric measurements in the whole spectral range. 
Since our sample selection was based on the {WIDE-S} 90~$\mu$m AKARI band, each galaxy  has at least one measurement in the FIR, at 90~$\mu$m. 
Additionally, 95\% of sources are detected in the MIR, all galaxies have optical information, and half of them were also detected in the UV. 
The typical spectral coverage of a source from our sample is shown in Fig.~\ref{6M}.
After this selection, 95 galaxies from the initial sample of 173 galaxies with a known spectroscopic redshift remained {for} the subsequent analysis. 
All the available measurements from the \mbox{ADF--S} database were used for the SED fitting. 

\begin{figure}[t]
		\centerline{\includegraphics[scale=0.5,clip]{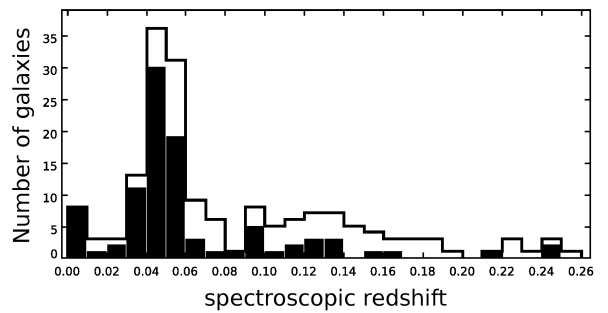}}
		\caption[]{The spectroscopic redshift distribution N($z$) in 0.01 $z$ bins.
		The open histogram corresponds to the distribution of 173 galaxies with known spectroscopic redshifts in our  sample. 
		{ The only o}ne object with a redshift higher than 0.3 is not shown here (quasar HE 0435-5304, located at z=1.232).		
		 {The filled histogram} corresponds to the 95 galaxies used for Spectral Energy Distribution fitting.
		}
		\label{SPEC}
	\end{figure}

The redshift distribution $N(z)$ of the selected sample with spectroscopic redshifts is presented in Fig.~\ref{SPEC}. 

\section{Photometric redshifts}
\label{photoredshifts}

The information about spectroscopic redshifts is available for 173 galaxies among 545 galaxies identified by {Ma{\l}ek} et~al. (2010). 
It implies that more than 68\% of sources in our sample have an identification in public catalogues as galaxies with photometric data but no redshift.  
In order to analyse properties of all the identified galaxies, we decided to estimate the photometric redshifts ($z_{\mathrm{phot}}$) for all galaxies with at least 6 measurements in the whole spectral range (127 galaxies fulfill this condition). 

We performed a test of the accuracy of the photometric redshift {for} the sample of 95 galaxies with known $z_{\mathrm{spectro}}$ and at least 6 photometric measurements { in} the whole spectral range (see Fig.~1). 
As representative parameters describing the accuracy of our method we used the percentage of successfully estimated redshifts, and the percentage of catastrophic errors (hereafter CE), which meet { the} condition:
\begin{equation}
\rm{ {\rm CE} := \frac{|z_\mathrm{spectro}-z_\mathrm{photo}|}{(1+z_\mathrm{spectro})}>0.15,} 
\end{equation}
following {Ilbert}~et~al.~(2006). 
The fraction of CE (hereafter $\eta$) is defined as the ratio of galaxies for which CE occurred and all the galaxies in the sample.  
The redshift accuracy $\rm{\sigma_{\Delta z/(1+z)}}$ was measured using the normalized median absolute deviation:
\begin{equation}
 \rm{\sigma_{\Delta z/(1+z)}=1.48\cdot median \frac{|\Delta z|}{1+z_{spectro}},}
 \end{equation}
where $\Delta z$ is the difference between spectroscopic and estimated photometric redshift. 

To estimate the photometric redshifts we used two different codes:
\subsubsection*{-- Le~PHARE:}

\textit{Photometric Analysis for Redshift Estimate} version 2.2 (Le~PHARE\footnote{http://www.cfht.hawaii.edu/$\sim$arnouts/LEPHARE/lephare.html}; {Arnouts} et~al., 1999, {Ilbert} et~al., 2006) code. 
We tested all the available libraries{, one by one,} of galaxy SEDs included in the Le~PHARE distribution. 
Using the sources with  spectroscopic redshifts available as a test sample, we checked the performance of all the available libraries: 
the percentage of sources for which the photometric redshift estimation could be performed successfully, and the scatter between estimated photometric and ''real'' spectroscopic values of redshifts.
The best results were obtained for the Rieke library, which includes eleven SEDs of luminous star forming galaxies constructed by Rieke et al. (2009). 
This is not very surprising, since it is expected that most of our galaxies belong to this class of galaxies.

\subsubsection*{-- CIGALE:}
Considering that a significant part of our data is in the FIR, which is not well covered by the templates included in the Le~PHARE distribution, we decided to test a different approach, i.e. 
to use the \textit{Code Investigating GALaxy Emission}\footnote{http://cigale.oamp.fr/} (CIGALE; {Noll} et~al., 2009) SED fitting program as a tool for the estimation of photometric redshifts. 
CIGALE was not developed as a tool for the estimation of $z_{\mathrm{phot}}$ but since it uses a large number of models covering the wide spectral range, including IR and FIR, it may be expected to provide a better $z_{\mathrm{phot}}$ for our FIR-selected sample than the software that mainly uses optical to NIR data. 

In the case of CIGALE, the values of parameters defined above were: $\eta$=9.47\% and $\rm{\sigma_{\Delta z/(1+z)}}$=0.06. 
We found that less than 10\% of $z_{\mathrm{CIGALE}}$ { suffer} from CE. 
CE occurred in the case of nine galaxies, and the mean value of $|z_\mathrm{spectro}-z_\mathrm{CIGALE}|/(1+z_\mathrm{spectro})$ for the CE subsample was equal to 0.22 $\pm$ 0.06, with minimum and maximum values of CE equal to 0.16 and 0.34, respectively.

Performing the same test with Le~PHARE we obtained $\eta$=14.73\% and $\rm{\sigma_{\Delta z/(1+z)}}$=0.05. 
We were not able to estimate photometric redshifts for 11 galaxies, and, additionally, CE occurred { for} three galaxies. 
The mean value of $|z_\mathrm{spectro}-z_\mathrm{phot}|/(1+z_\mathrm{spectro})$ for the CE subsample was equal to 0.27 $\pm$ 0.02 (with minimum and maximum values of CE equal to 0.26 and 0.28, respectively). 
{Thus}, we conclude that the deviation from the real value for the successful measurements was lower in the case of Le~PHARE; however, the percentage of successful measurements was higher when CIGALE was used. Thus, Le~PHARE provides higher accuracy while CIGALE assures a higher success rate for our FIR-selected sample. 
We have also checked that the amplitude of CEs (both mean and median values) was smaller in the case of CIGALE. 
For example, the minimum value of CE for CIGALE equals 0.16 (two galaxies), and it is very close to the CEs boundary limit.

This result is most likely related to our sample selection (which consists of galaxies bright at 90~$\mu$m), and its spectral coverage, in particular a small amount of optical and MIR measurements which are needed by Le~PHARE to determine the galaxy properties (e.g. Balmer break) properly.   
The possible explanation of a better performance of CIGALE is the limited number of FIR templates used by Le~PHARE. 
Additionally, according to Ilbert~et~al. (2006), Le~PHARE has the best performance in the redshift range 0.2$\leq$z$\rm{_{spectro}} \leq$1.5, while our spectroscopic range (0$\leq$z$\rm{_{spectro}} \leq$0.25) lies outside this redshift range. 
For galaxies with spectroscopic redshifts lower than 0.2 Ilbert~et~al.~(2006) have found a dramatic increase of the fraction of CE, caused by the mismatch between the Balmer break and the intergalactic Lyman-alpha forest depression at $\lambda <$ 1216 \AA{}. 
Yet another reason for the lower percentage of redshifts measured successfully by Le~Phare are the spectral types of galaxies which dominate in our sample: photometric redshifts of actively star forming galaxies are less reliable (the fraction of CE increases by a factor of $\sim$ 5 from the elliptical to the starburst galaxies for a sample used by Ilbert~et~al., 2006). 

Consequently, for the subsequent analysis we decided to use $z_{\mathrm{CIGALE}}$ for sources without a known $z_\mathrm{spectro}$.
A more detailed comparison of the estimation of photometric redshifts by Le~PHARE and CIGALE can be found in {Ma{\l}ek} et~al. (2013).
  
\section{SED fitting}  
\label{fitting}
\begin{figure}[t]
	\centerline{\includegraphics[scale=0.5,clip]{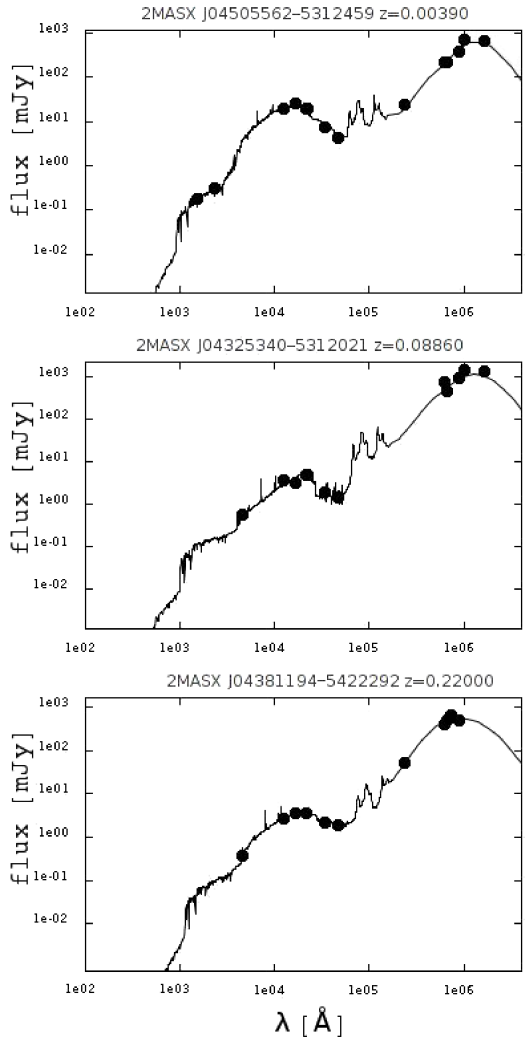}}
	\caption[]{ Three examples of the best-fit models for the normal galaxy (upper panel), LIRG (middle panel) and ULIRG (bottom panel) sources.  
	Solid lines correspond to the best model obtained from CIGALE code, and the full black circles - observed data used for SED fitting.}
	\label{3sedy}
\end{figure}
To study the physical parameters of the \mbox{ADF--S} sources, we selected galaxies with a known $z_{\mathrm{spectro}}$ or $z_{\mathrm{CIGALE}}$, and with the highest quality photometry available. 
The main selection criteria were to have redshift information, together with at least six measurements in the whole spectral range. 
As a result, we use 222 galaxies: 95 sources with $z_{\mathrm{spectro}}$ and 127 galaxies with $z_{\mathrm{CIGALE}}$.
All available photometric measurements for these galaxies were used for the SED fitting with CIGALE. 

CIGALE uses models describing the emission from a galaxy in the wavelength range from the rest-frame far-UV to the {rest--frame} far--IR (Noll et~al., 2011).
The code  { derives} physical parameters of galaxies { by} fitting their spectral energy distributions (SEDs) to SEDs based on models and templates.
CIGALE takes into account both the dust UV attenuation and IR emission. 
Based on possible values for each physical parameter related to star formation history, dust attenuation, and dust emission, CIGALE computes all possible spectra and derives mean fluxes in the observed filters. 
For each galaxy, the best value for each parameter, as well as the best fitted model, is found using a Bayesian-like statistical analysis (Roehlly et~al., 2012). 

Models of  stellar emission are given either by Maraston (2005) or Fioc \& {Rocca-Volmerange} (1997).
The absorption and scattering of stellar light by dust, the so-called attenuation curves for galaxies, are given by Calzetti et~al. (2000).
Dust emission is characterized by a power-law model proposed by Dale and Helou (2002), with the slope $\mathrm{\alpha_{SED}}$ of the relation between the dust mass and the heating intensity. 
This is the only dust emission model included in the newest CIGALE distribution (CIGALE version {2013/01/02}). 

To reconstruct more accurately Star Formation Rates (SFRs), CIGALE uses the single stellar population of Maraston (2005). 
For the old stellar population, CIGALE calculates old stellar population history (SFH$_1$) with exponential decrease:
\begin{equation}
 \rm{SFR_{1}=\frac{M_{gal}}{\tau_{1}(e^{1/\tau_{1}}-1)}}
\end{equation}
In our work we have adopted the box model for the young stellar population history (SFH$_2$), with constant star formation over a limited period of time, starting from 0.0025 to 1 Gyr ago (parameter $t_2$ in Table~\ref{cigale_parameters}). 
In the case of a box model, the SFH$_{2}$ is computed as the galaxy mass divided by its age.
Thus, CIGALE gives a total value of logSFR, defined as:
\begin{equation}
 \rm{SFR=(1-f_{ySP})SFR_{1} + f_{ySP}SFR_{2},}
\end{equation}
where $\rm{f_{ySP}}$ is the fraction of the young stellar population.

The list of input parameters of CIGALE is shown in Table~\ref{cigale_parameters}.

\begin{table*}[t]
\small
\renewcommand{\arraystretch}{1.2}
\vspace{-.3cm}
\caption{List of the input parameters of CIGALE, based mostly on Buat et~al. (2011).}
\vspace{-.1cm}
\begin{center}
\begin{tabular}{p{6cm} | p{1.5cm} | p{6.5cm}} \hline
\label{cigale_parameters}
 &  Symbol & Range \\ \hline \hline
Star formation history & &  \\  \hline
metallicities (solar metalicity) & Z & 0.2        \\                    
$\tau$ of old stellar population models in Gyr  & $\mathrm{\tau_{1}}$ & 1. 3. 10. \\
ages of old stellar population models in Gyr  & $\mathrm{t_{1}}$ & 13\\
ages of young stellar population models in Gyr  &  $\mathrm{t_{2}}$ & 0.025 0.05 0.1 0.3 1.0\\
fraction of young stellar population &  $\mathrm{f_{yS P}}$ & 0.001 0.01 0.1 0.999 \\ \hline
{Dust attenuation }& & \\ \hline
slope correction of the Calzetti law &  $\delta$ & -0.3 -0.2 -0.1 0. 0.1 0.2\\
V-band attenuation for the young stellar population  & $\mathrm{A_{V,yS P}}$ & 0.15 0.30 0.45 0.60 0.75 0.90 1.05 1.2 1.35 1.5 1.65 1.8 1.95 2.10 \\
Reduction factor of the dust attenuation at 5500\AA{} for old stellar population models  & $\mathrm{f_{att}}$ & 0.0 0.5 1.0  \\ \hline
{Dust emission} & & \\ \hline
IR power-law slope &  $\alpha_{\rm{SED}}$&  0.5 1.0 1.5 1.75 2.0 2.25 2.5 4.0  \\    
$\mathrm{L_{dust}}$ fraction of AGN model & $\mathrm{f_{AGN}}$ & 0. 0.1 0.25 \\    \hline      
\end{tabular}
\end{center}
\end{table*}

The reliability of the retrieved parameters for galaxies with known spectroscopic redshifts was checked using the mock catalogue of artificial galaxies ({Ma{\l}ek} et~al., 2013). 
{ The comparison between the results from the mock and real catalogues shows that CIGALE gives a very good estimation of stellar masses, star formation rates, ages sensitivity D4000 index, dust attenuations and dust emissions, bolometric and dust luminosities (with values of the linear Pearson moment correlation coefficient, \textit{r}, higher than 0.8).
The accuracy of the relation between the dust mass and the heating intensity, $\alpha_{\rm{SED}}$, is estimated with a lower efficiency (\textit{r}=0.55). }

\section{Physical properties of \mbox{ADF--S} sample}
\label{result}
We restrict the further analysis to the SEDs with a minimum value of $\chi^2$ lower than four. 

\begin{figure}[t]\centering
    {\includegraphics[scale=0.5,clip]{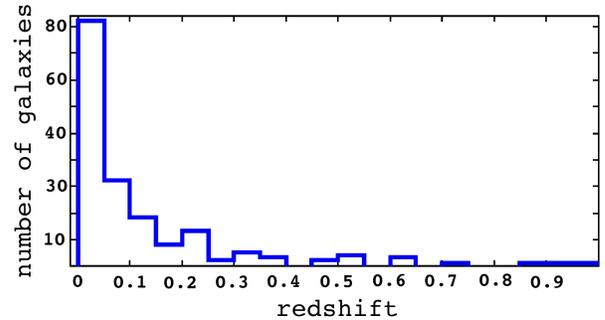}}
        \caption{The distribution of redshifts in our { final} sample (both spectroscopic and estimated by CIGALE).}
     \label{TotalZ}
\end{figure}

This condition was met by 186 galaxies (73 galaxies with $z_{\mathrm{spectro}}$ and 113 sources with an estimated $z_{\mathrm{CIGALE}}$). 
Consequently, we assume that SEDs were successfully fitted only for this final sample. 
The redshift distribution of this sample is shown in Fig.~\ref{TotalZ}.
 Examples  of the best fit models obtained from CIGALE are given in Fig.~\ref{3sedy}. 

\begin{figure*}[t]\centering
  	\includegraphics[scale=0.33,clip]{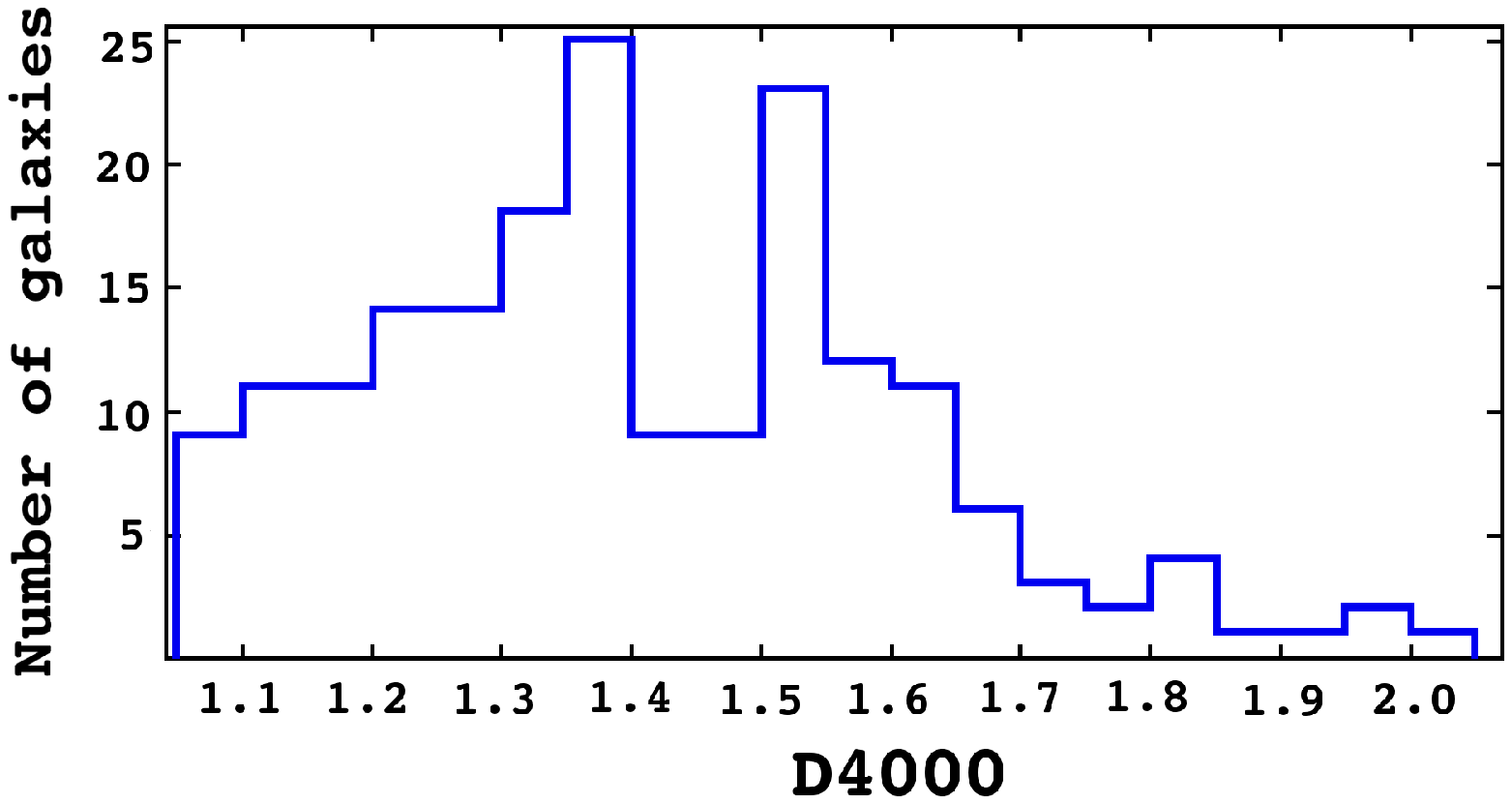}\includegraphics[scale=0.33,clip]{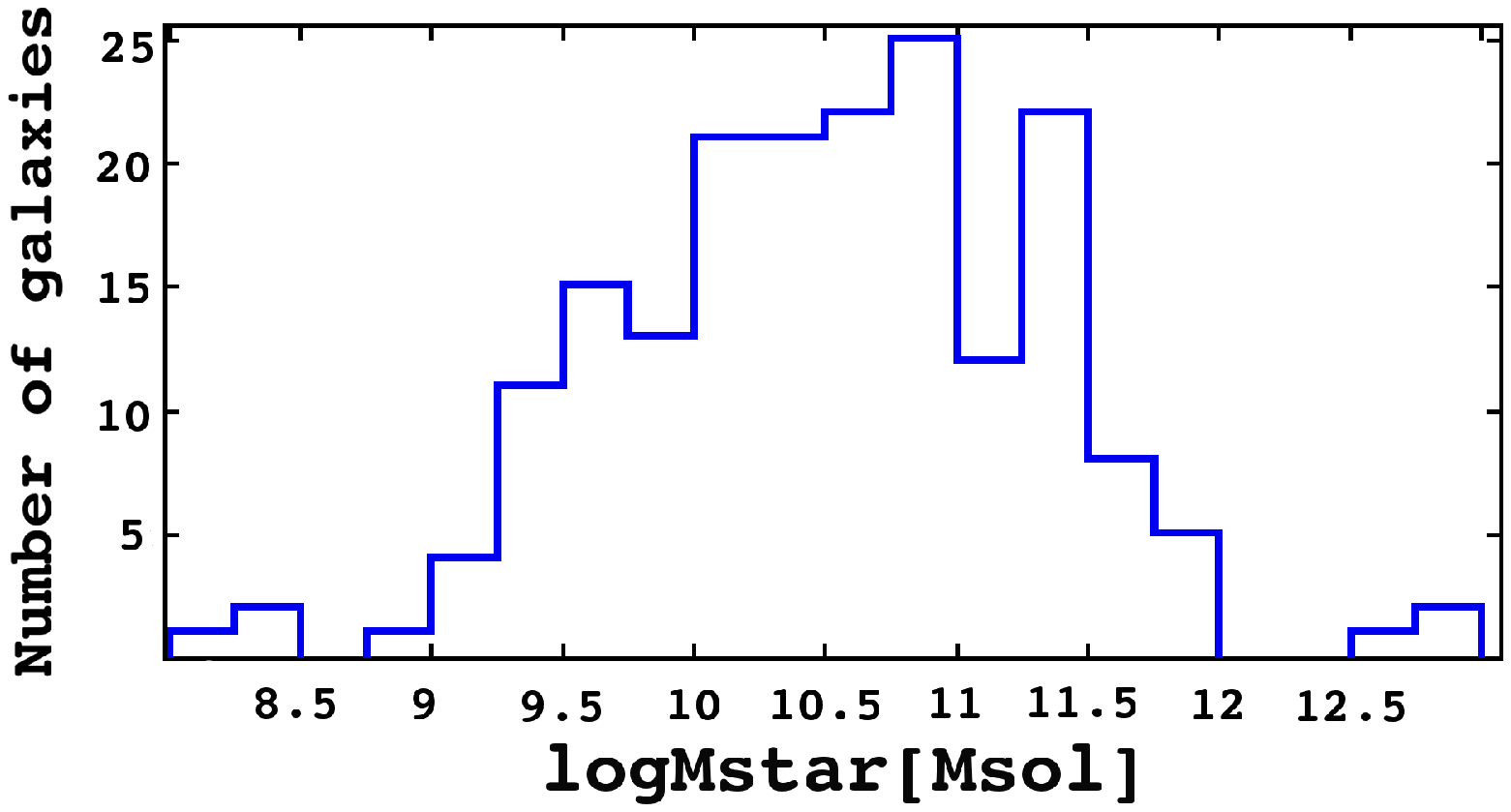}\includegraphics[scale=0.33,clip]{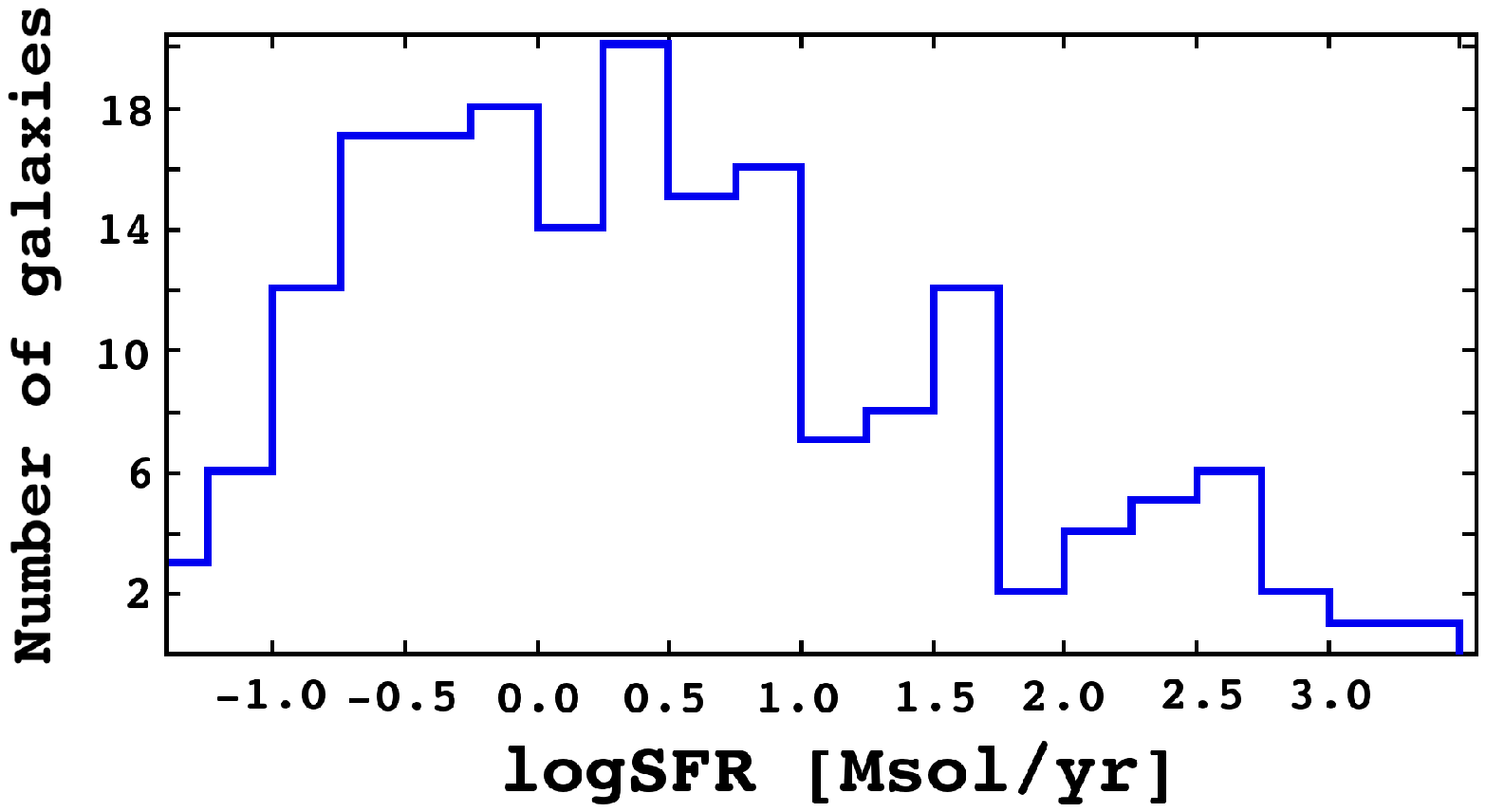}
	\includegraphics[scale=0.33,clip]{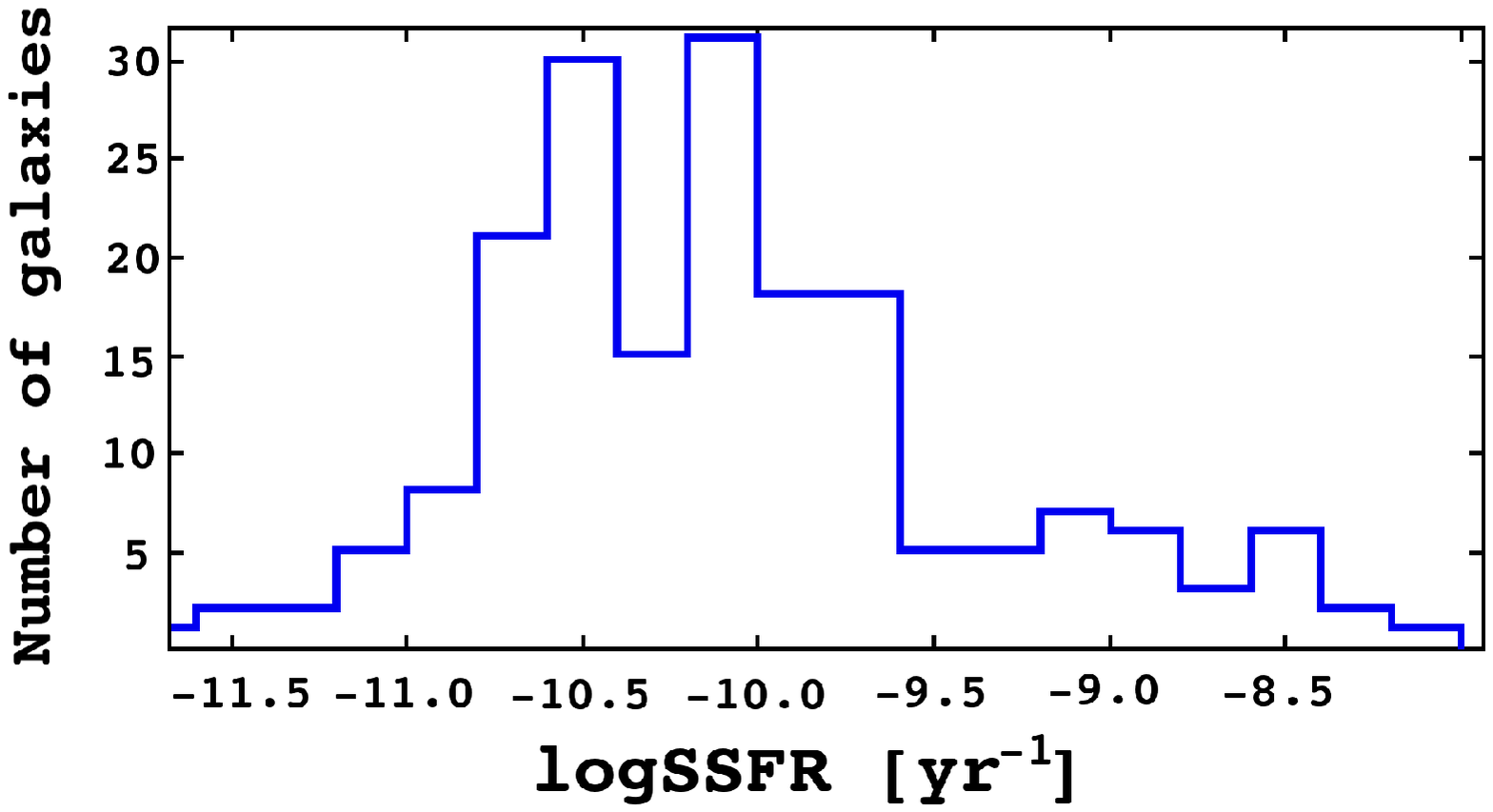}\includegraphics[scale=0.33,clip]{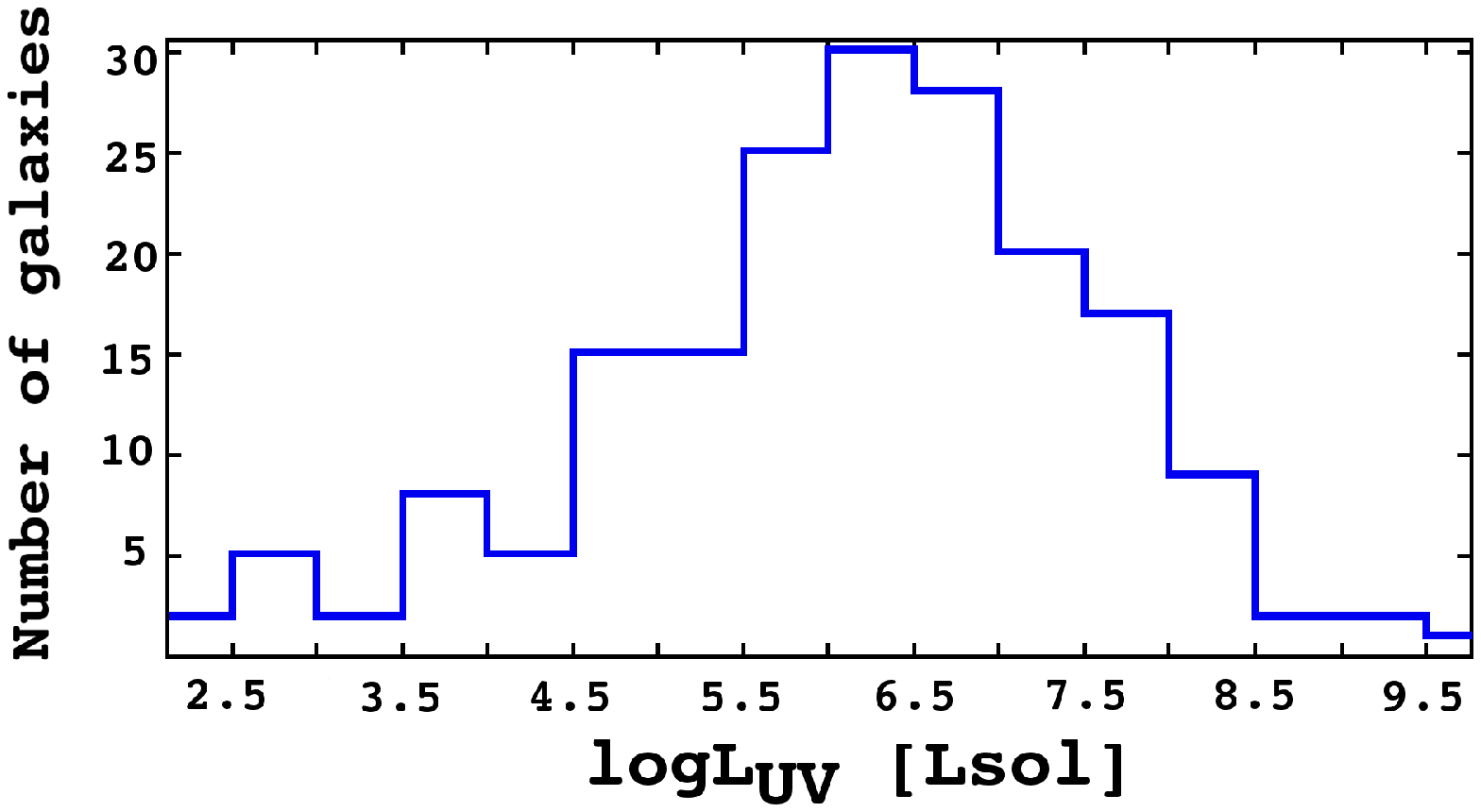}\includegraphics[scale=0.33,clip]{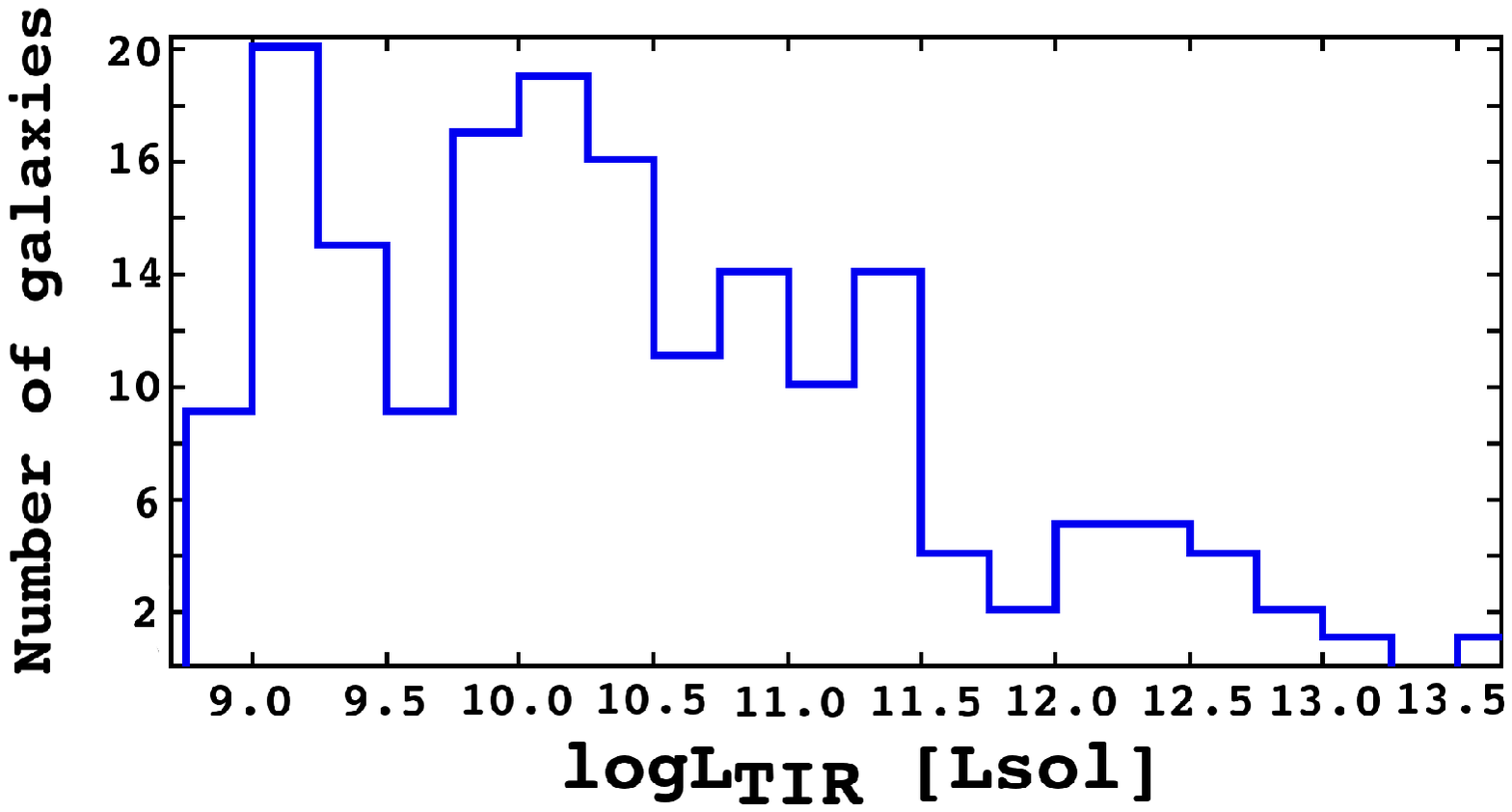}
         \caption{Distribution of the Bayesian estimates of the output parameters: D4000, $\mathrm{M_{star}}$,  and  $\mathrm{logSFR}$. 
         SSFR values were calculated as the ratio of SFR and $\mathrm{M_{star}}$.
         $\mathrm{L_{UV}}$, and $\mathrm{L_{TIR}}$ were calculated as an integral value of SEDs from 1480$\cdot$(1+z) to 1520$\cdot$(1+z) \AA{}, and from 8$\cdot$(1+z) $\mu$m to 1$\cdot$(1+z)~mm, respectively.}
\label{comparison_param_both_samples}     
  \end{figure*}

The distribution of the main parameters (estimated with the Bayesian analysis for 186 galaxies in our final sample) is plotted in Fig.~\ref{comparison_param_both_samples}.
We found that galaxies in our sample are typically very massive, with a mean value  of $\mathrm{M_{star}} =  10.48  \pm 0.19\cdot10 ^ {10}\mbox{ }[\rm{M_\odot}]$.
Moreover, these galaxies are rather luminous $\mathrm{L_{bol}} = 10.81 \pm 0.93 \cdot 10 ^ {10}\mbox{ }[\rm{L_\odot}]$,  and also  their dust luminosity is high $\mathrm{L_{dust}} = 10.38 \pm 1.01 \cdot 10 ^ {10}\mbox{ }[\rm{L_\odot}]$, but without a precisely defined maximum. 
A median value of the star formation rate parameter, SFR, is equal to 2.22 $[\rm{M_\odot yr^{-1}}]$. 
The estimated value of the heating intensity $\alpha_\mathrm{SED}$ (Dale \& Helou, 2002) implies that the vast majority of analysed galaxies (85.48\%) belong to a normal, star-forming galaxy population, with median the value of $\alpha_\mathrm{SED}$ equal to { 2.01}. 
The median value of the $\mathrm{A_V}$ parameter, describing the effective dust attenuation for the stellar population at a wavelength equal to 5500 \AA{} is 0.47 [mag], and the median value for the attenuation in the FUV (at 1500 \AA{}, $\mathrm{A_{FUV}}$) is 1.87 [mag].
The parameter $\mathrm{A_{V,ySP}}$, which describes the V-band attenuation for the young stellar population model, spreads almost { across} the entire range of input parameters from 0.15 to 2.19, with the median value 0.97 [mag].
  
\section{Average Spectral Energy Distributions}
\label{resultAVSEDS}

\begin{figure}[t]\centering
    {\includegraphics[scale=0.5,clip]{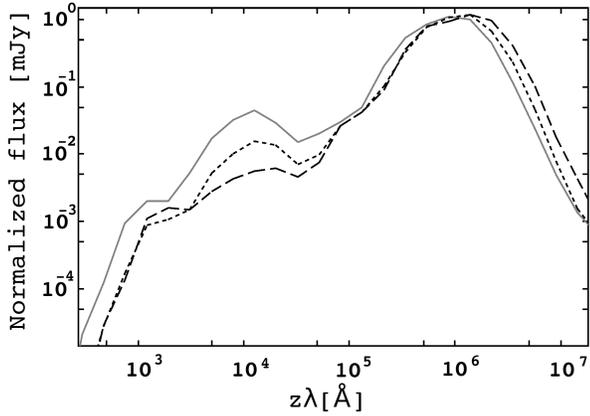}}
        \caption{The average SEDs, normalized at 90 $\mu$m, of ULIRGs (dashed line), LIRGs (dotted line) and the remaining  galaxies (solid line).
        SEDs were shifted to the rest frame.}
     \label{Mean_sed}
\end{figure}
  
Using the CIGALE output, we created average SEDs from our 186  galaxies. 
First, we normalized all SEDs at a rest frame 90 $\mu$m; then we divided them into 3  broad categories: Ultraluminous Infrared Galaxies (ULIRGs), Luminous Infrared Galaxies (LIRGs), and the remaining galaxies.

Following {Sanders} \& {Mirabel (1996), we define ULIRGs as galaxies with a very high IR luminosity, $\mathrm{L_{TIR}>10^{12} L_{\odot}}$, where  $\mathrm{L_{TIR}}$ is the total mid- and far-infrared luminosity calculated in the range between 8(1+z)~$\rm{\mu}$m and $1(1+z)$~mm. 
Sources with less extreme, but still high, IR luminosities $\mathrm{10^{11}L_{\odot} < L_\mathrm{TIR}< 10^{12} L_\mathrm{\odot}}$ are classified as LIRGs.
In our sample, we found 18 ULIRGs (9.7\% of analysed \mbox{ADF--S} sources) and 30 LIRGs (16.1\% of the total number of sources).  

Average SEDs for ULIRGs, LIRGs and the remaining galaxies are plotted in Fig.~\ref{Mean_sed}. 

The ratio of bolometric to total mid- and far-infrared luminosity (the integrated luminosities calculated from CIGALE) is higher for the ULIRGs and LIRGs than for the normal galaxies. 
In the case of the average SEDs, the $\rm{L_{bol}/L_{TIR}}$ ratio is equal to 0.73 $\pm$ 0.16 for { the} ULIRGs, 0.55 $\pm$ 0.16 for the LIRGs,  and 0.39 $\pm$ 0.22 for the remaining galaxies. 
Both ULIRGs and LIRGs in our sample contain dust which is cooler than the dust in the remaining galaxies, which can be seen as a shift of the maximal values of the dust components towards longer wavelengths. 
The brighter the sample is in the IR, the more shifted is the dust peak towards the longer  wavelengths. 
In our sample, the maximum of the dust peak in the spectra normalized to 90~$\mu$m is located at 1.38$\cdot10^6$\AA{}, 1.21$\cdot10^6$\AA{}, and 8.65$\cdot10^5$\AA{} for the ULIRGs, LIRGs and normal galaxies, respectively. 

The median redshift for ULIRGs in our sample is equal to 0.54.
The median redshift for LIRGs, and the remaining galaxies was found to be 0.2, and 0.04, respectively. 
This difference in redshifts is a selection effect, related to the fact that the primary detection limit is in the FIR. 
The two parameters SFR and L$\rm{_{TIR}}$ might be correlated because  both of them depend on the galaxy mass, but they are estimated independently.
For the ULIRGs, the logSFR is very high, with a mean value 2.59 $\pm$ 0.32 $\mathrm{[M_{\odot}yr^{-1}}]$. 
The SFR for the LIRGs is more than ten times lower (the mean logSFR is equal to 1.47 $\pm$ 0.28 $\mathrm{[M_{\odot}yr^{-1}}]$). 

The SFR for the normal galaxies is much lower: we found the logSFR on the level of -0.06 $\pm$ 0.62 $[\mathrm{M_{\odot}yr^{-1}]}$.
Comparing our results to the sample of normal, nearby galaxies from SINGS (Kennicutt et al., 2003) we found a similar range of SFR: 0-12 $\mathrm{[M_{\odot}yr^{-1}}]$ for the ADF-S, and 0-15 $\mathrm{[M_{\odot}yr^{-1}}]$ for the SINGS sample.  }

Comparing the results obtained by U~et al. (2012) for a sample of 53 LIRGs and 11 ULIRGs in the similar redshift range (z between 0.012 and 0.083) from the Great Observatories All-sky LIRG Survey (GOALS), we found a very good agreement of SFR for LIRGs (logSFR$\rm{_{LIRGs}}$ = 1.57 $\pm$ 0.19 $\mathrm{[M_{\odot}yr^{-1}]}$). 
The mean value of logSFR$\rm{_{ULIRGs}}$ in the case of our \mbox{ADF--S} sample is higher than the one found by U et al. (2012) and equals logSFR$\rm{_{ULIRGs}}$ = 2.25 $\pm$ 0.16 $\mathrm{M_{\odot}yr^{-1}}$, but is still consistent within the error bars. 

The LIRGs in the ADF-S sample have a mean logM$\rm{_{star}}$ at 11.15 $\pm$ 0.49 $\mathrm{[M_{\odot}]}$. 
This value is { slightly} higher than logM$\rm{_{star}}$ for LIRGs reported by U et al. 2012 (10.75 $\pm$ 0.39 $\mathrm{[M_{\odot}]}$),
Giovannoli et al. 2011 (LIRGs sample, with logM$\rm{_{star}}$ between 10 and 12, and with a peak at 10.8), and Melbourne et al. 2008 (logM$\rm{_{star}}$ $\sim$ 10.5, based on a set 15 LIRGs, at redshift $\sim$ 0.8). 
Even though our result shows the highest value of logM$\rm{_{star}}$, it  is still consistent with the values listed above within the error bars. 
Stellar masses for the ULIRG sample have a mean logM$\rm{_{star}}$ equal to 11.43 $\pm$ 0.31 $\mathrm{[M_{\odot}]}$. 
The mean stellar masses for 11 ULIRGs presented by U et al. (2012) were calculated as logM$\rm{_{star}}$=11.00 $\pm$ 0.40 $\mathrm{[M_{\odot}]}$. 
The sample of ULIRGs analysed by Howell et al. (2010) is characterized by a  mean stellar mass logM$\rm{_{star}}$=11.24 $\pm$ 0.25 $\mathrm{[M_{\odot}]}$. 
We conclude that the (U)LIRGs in our \mbox{ADF-S} sample are slightly more massive than those in the samples used in previous works, but still all the samples are statistically consistent.

\begin{figure}[t]
		\centerline{\includegraphics[scale=0.4,clip]{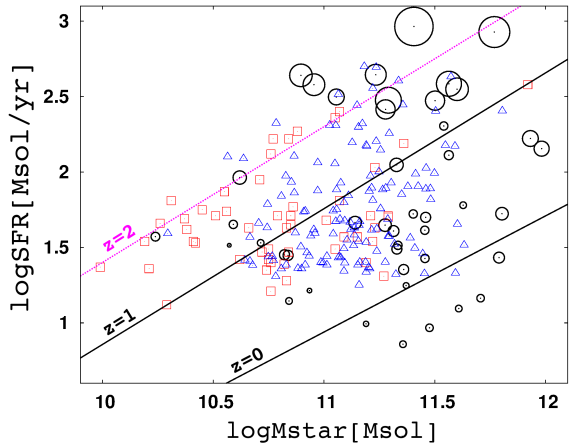}}
		\caption[]{The SFR vs stellar mass relation  for \mbox{ADF--S} (U)LIRG sample (black circles). 
		The variable size of points represents the difference in redshift (increasing size from $z$ equal to 0.09 to 0.98).
		We overplot the Howell et al., (2010) GOALS sample (open blue triangles), and U et al. (2012) (U)LIRG sample (red open squares). 
		The two solid black lines correspond to the SFR-Mstar observation from Elbaz et al. (2007). 
		The { violet} dashed line indicates the SFR - stellar mass relation for star-forming galaxies at z$\sim$2 defined by Daddi et al. (2007).		}
		\label{SFRMs}
	\end{figure}

Figure~\ref{SFRMs} shows the relation between the SFRs and stellar masses for (U)LIRGs in our \mbox{ADF--S} catalogue. 
The mean redshift of the (U)LIRG sample is equal to 0.34 (minimum and maximum redshifts are equal to 0.09 and 0.98, respectively - then, the redshift distribution is { quite} broad). 
We compared our results with observations at redshifts 0 and 1 (Elbaz et al., 2007), and 2 (Daddi, et al., 2007), and also with other data mentioned above: a sample of nearby (z$<$0.032) (U)LIRGs (U et al., 2012), and
a sample of LIRGs observed in the GOALS  by GALEX and the Spitzer Space Telescope (Howell et al., 2010). 
We found a rather flat distribution of the SFR {parameter in the stellar mass space} (similar to Giovannoli et al., 2011, { who} found for a sample of LIRGs in the Extended Chandra Deep Field South at z = 0.7, selected at 24 $\mu$m by Spitzer). 
{ Our results, shown in Fig.~\ref{SFRMs}, confirm a strong correlation between SFR and redshift.} 

Based on the physical properties obtained from SED fitting, we computed the  Specific Star Formation Rates (SSFR [yr$^{-1}$]; defined as the ratio of SFR and stellar mass) for the \mbox{ADF--S} sample. 
The SSFR rate is commonly used to analyse star formation history. 
For ULIRGs, LIRGs, and the rest of our sample, the logarithmic values of the SSFR parameter are equal to -9.00 $\pm$ 0.55, -9.68 $\pm$ 0.59, and -10.28 $\pm$ 0.57, respectively. 
{Howell} et al. (2010), and U.~et al. (2012) presented the mean and median values of the SSFR for (U)LIRGs sample, without an additional separation. 
The median value of SSFR for galaxies in the GOALS field ({Howell} et al., 2010) is equal to -9.41 [yr$^{-1}$], while U et al. (2012) obtained the value of -9.17 [yr$^{-1}$]. 
The same parameter computed for LIRGs and ULIRGs together in our \mbox{ADF--S} sample is equal to -9.51  [yr$^{-1}$]. 
We conclude that our results are consistent with the results mentioned in all the other works usedconsidered for comparison. 

However, the dust emission power-law model, given by Dale and Helou (2002) implemented in CIGALE may not be efficient enough to describe the properties of the (U)LIRG sample well.  
In the future we plan to apply other models (e.g. Siebenmorgen \& Kr\"{u}gel, 2007, Chary \& Elbaz, 2001, Casey, 2012) {to} the same (U)LIRGs sample.

\section{Conclusions}
\label{conclsection}

\begin{itemize}
 \item For our analysis we used an AKARI \mbox{ADF--S} catalogue ({Ma{\l}ek} et~al., 2010), with additional information about spectroscopic redshifts (Ma{\l}ek et~al., 2013)
 \item The CIGALE (Noll et al., 2009), program for fitting SEDs, was used for the first time to estimate photometric redshifts for galaxies without known spectroscopic redshifts (127 galaxies), and then to fit Spectral Energy Distribution models to our whole sample.
 We conclude that although CIGALE was not designed as a tool for the estimation of $z_\mathrm{phot}$, in the case of the \mbox{ADF--S} galaxies it is more efficient than the standard code (Le~PHARE), based on the UV and optical data. 
This observation might be worth taking into account for IR-selected galaxies  in the future analysis. 
 \item We used CIGALE  for the redshift estimation for 127 galaxies with at least 6 photometric measurements in the whole spectral range, from UV to FIR. 
 A satisf{actory} value of the $\chi^2$ parameter for the fit, lower than 4, was obtained for 186 galaxies. 
 \item Based on the physical parameters obtained from SED fitting, we conclude that our FIR selected sample consists mostly of nearby, massive galaxies, bright in the IR, and active in SF (a similar conclusion was found by White et~al., 2012).
 The distribution of the SFR shows that our sample is characterized by a rather high star formation rate, with a the median value equal to 1.96 and 2.56 [$\mathrm{M_{\odot}yr^{-1}}$], for galaxies with spectroscopic and estimated photometric redshift, respectively,
  \item Average SEDs for ULIRGs, LIRGs and remaining galaxies from our sample, normalized at 90 $\mu$m  were created. 
  Almost 25\% of our sample are (U)LIRGs, rich in dust and active in star formation processes. 
  For these galaxies we noticed a significant shift in the peak wavelength of the dust emission in the FIR and a different ratio between luminosities in the optical and IR {parts of the} spectra.
 \end{itemize}

\acknowledgments{
We would like to thank both anonymous Reviewers for their very constructive comments and suggestions which helped to improve the quality of this paper. 
We thank Olivier Ilbert for useful discussions and kind help in using Le~PHARE. 

This work is based on observations with AKARI a JAXA project with the participation of ESA. 
This research has made use of the SIMBAD and NED databases.
KM, AP and AK were financed by the research grant of the Polish Ministry of Science N N203 512938. 
The collaboration between French and Polish participants was partially supported by the European Associated Laboratory Astrophysics Poland-France HECOLS. 
This research was partially supported by the project POLISH-SWISS ASTRO PROJECT co-financed by a grant from Switzerland through the Swiss Contribution to the enlarged European Union.
KM has been supported from the Japan Society for the Promotion of Science (JSPS) Postdoctoral Fellowship for Foreign Researchers, P11802.
TTT has been supported by the Grant-in-Aid for the Scientific Research Fund (20740105, 23340046, and 24111707) and for the Global COE Program Request for Fundamental Principles in the Universe: from Particles to the Solar System and the Cosmos commissioned by the Ministry of Education, Culture, Sports, Science and Technology (MEXT) of Japan.
VB and DB have been supported by the Centre National des Etudes Spatiales (CNES) and the Programme National Galaxies (PNG). 
MM acknowledges support from NASA grants NNX08AU59G and NNX09AM45G for analysis of the GALEX data in the Akari Deep Fields. 
}

\email{K. Ma{\l}ek (e-mail: malek.kasia@nagoya-u.jp)}
\label{finalpage}
\lastpagesettings
\end{document}